# The covariant electromagnetic Casimir effect for real conducting spherical shells

**H. Razmi** [(1)] and **M. Abtahi** [(2)]

Department of Physics, The University of Qom, Qom 3716146611, I. R. Iran.
(1) razmi@qom.ac.ir & razmiha@hotmail.com  (2) mahdiyeh.abtahi@gmail.com

**Abstract**

Using the covariant electromagnetic Casimir effect (previously introduced for real conducting cylindrical shells [1]), the Casimir force experienced by a spherical shell, under Dirichlet boundary condition, is calculated. The renormalization procedure is based on the plasma cut-off frequency for real conductors. The real case of a gold (silver) sphere is considered and the corresponding electromagnetic Casimir force is computed. In the covariant approach, there isn't any decomposition of fields to TE and TM modes; thus, we do not need to consider the Neumann boundary condition in parallel to the Dirichlet problem and then add their corresponding results.

# Introduction

Although calculation of the Casimir effect with spherical boundaries is one of the first important generalizations of the primary work studied by Casimir for two parallel conducting plates [2], the corresponding mathematical arguments and manipulations are more complicated and even controversial. A few authors (including Casimir [3]) have argued that the Casimir force for a conducting spherical shell is attractive, similar to the case of two parallel plates (e.g. [4]). Most of the currently known calculations, including the original work introduced by Boyer in 1968 [5], reason on this fact that the Casimir effect for this particular geometry is repulsive [6–8]. Application of the Casimir effect in the spherical geometry is also controversial. For example, considering the phenomenon of sonoluminescence [9–11], some scientists (including Julian Schwinger) have argued about the relevance of this phenomenon to the Casimir effect and some others have tried to show that sonoluminescence is irrelevant to the Casimir effect [12–31]. All of these researches, with their own special technique (e.g. the Green function or mode by mode summation method) of computing the Casimir energy/force, deal with non-covariant formulation of the electromagnetic field. Indeed, in the already known calculations of the electromagnetic Casimir effect, the quantum vacuum fluctuations corresponding to the non-covariant quantities $\vec{E}$ and $\vec{B}$ (the electric and magnetic fields separately) are considered. As we know, it is guessed that the Casimir effect originates from the virtual particles (photons) of the quantum vacuum state. These quantum field theoretical 'virtual' particles do not appear in a non-covariant quantum theory of the electric and magnetic fields; but, these vacuum particles are quanta of the covariant electromagnetic field tensor $F^{\mu\nu}$ based on the quantization of four-vector potential $A^{\mu}$. To have a better understanding of the quantum vacuum and the Casimir effect both conceptually and technically, one should work with the covariant quantization of the electromagnetic field. Here, we want to find the Casimir force experienced by a real conducting spherical shell using covariant formulation of the electromagnetic field. Our calculation is based on a cut-off frequency (the plasma frequency of the metal of which the spherical shell is made) regularization. At low frequencies, all the metals/conductors have a real and frequency independent conductivity; but, at frequencies higher than the plasma frequency, the electromagnetic field – here, the virtual photons causing the Casimir effect – 'see' the conducting spherical shell as transparent [32]. This means that for virtual photons with frequencies higher than the plasma frequency there is no boundary; in other words, there is no distinction between 'in' and 'out' of the shell and thus there is no Casimir effect. Only the photons whose frequencies are lower than the plasma frequency of the shell contribute to the Casimir effect. In computing the Casimir force, we restrict the upper bound of the integrals with plasma frequency cut-off. We should mention that although plasma frequency cut-off has been considered in the study of dielectrics (e.g. [33]), here we want to apply it to real conducting spherical shells. A number of technical methods, such as a summation of modes [34], dimensional regularization [35], and Green function method [6] have been employed to calculate the Casimir energy in different geometries. Among these well-known approaches, we use the Green function method based on covariant formulation of the electromagnetic field. Green function method [6] have been employed to calculate the Casimir energy in different geometries. Among these well-known approaches, we use the Green function method based on covariant formulation of the electromagnetic field.

# The electromagnetic Casimir effect: covariant formalism

In non-covariant formulation of the electromagnetic field, only the transverse radiation field is quantized. Since the decomposition of the field into transverse and longitudinal components is frame-dependent, this clearly hides the Lorentz-invariance of the theory. In the covariant formalism, all the components of the four-vector potential $A^\mu(x) = (\phi, \vec{A})$, as the dynamical variable under consideration, are quantized. The canonical stress tensor for the electromagnetic field is [36]

$$T^{\alpha\beta} = -g^{\alpha\mu} F_{\mu\lambda}(x) \partial^\beta A^\lambda(x) - g^{\alpha\beta} L_{EM} \tag{1}$$

where

$$L_{EM} = -\frac{1}{2} (\partial_\nu A_\mu(x))(\partial^\nu A^\mu(x)) \tag{2}$$

is the Fermi Lagrangian density*** for the free electromagnetic field and

$$F^{\mu\nu}(x) = \partial^\mu A^\nu(x) - \partial^\nu A^\mu(x) \tag{3}$$

is the electromagnetic field tensor whose components are frame-dependent (non-covariant) quantities $\vec{E}$ and $\vec{B}$.

With the substitution of (2) and (3) in (1), it is found

$$T^{\alpha\beta} = -g^{\alpha\mu}(\partial_\mu A_\lambda(x) \partial^\beta A^\lambda(x) - \partial_\lambda A_\mu(x) \partial^\beta A^\lambda(x)) + \frac{1}{2} g^{\alpha\beta}(\partial_\nu A_\mu(x) \partial^\nu A^\mu(x)) \tag{4}$$

This can be written as

$$T^{\alpha\beta} = \lim_{x \to x'}\Big[-g^{\alpha\mu}(\partial_\mu \partial'^\beta A_\lambda(x) A^\lambda(x')) - \partial_\lambda \partial'^\beta A_\mu(x) A^\lambda(x'))$$

$$+ \frac{1}{2} g^{\alpha\beta}(\partial_\nu \partial'^\nu A_\mu(x) A^\mu(x'))\Big] \tag{5}$$

Considering the operator form of the above relation and taking its vacuum to vacuum expectation value, it can be shown:

$$\langle 0|T^{\alpha\beta}|0\rangle = -i\hbar c \lim_{x \to x'}[-g^{\alpha\mu}(\partial_\mu \partial'^\beta g^\lambda_\lambda - \partial_\lambda \partial'^\beta g^\lambda_\mu) + \frac{1}{2} g^{\alpha\beta}(\partial_\nu \partial'^\nu g^\mu_\mu)]\Delta_F(x-x')$$

$$= -i\hbar c \lim_{x \to x'}[-g^{\alpha\mu}(4\partial_\mu \partial'^\beta - \partial_\mu \partial'^\beta) + \frac{1}{2} g^{\alpha\beta}(4\partial_\nu \partial'^\nu)]\Delta_F(x-x')$$

$$= i\hbar c \lim_{x \to x'}(3\partial^\alpha \partial'^\beta - 2g^{\alpha\beta} \partial_\nu \partial'^\nu)\Delta_F(x-x'), \tag{6}$$

in which we have used the following well-known relations:

$$\langle 0|T\{A^\mu(x) A^\nu(x')\}|0\rangle = i\hbar c D_F^{\mu\nu}(x-x'), \tag{7}$$

$$D_F^{\mu\nu}(x) = -g^{\mu\nu}\Delta_F(x) \qquad (8)$$

$$\partial_\mu \partial^\mu \Delta_F(x-x') = -\delta^{(4)}(x-x') \qquad (9)$$

$T\{A^\mu(x)A^\nu(x')\}$ is the time-ordered product of the field operators and $\Delta_F(x-x') = G(x,x') = G(\vec{x},t;\vec{x}',t')$ is the famous Feynman delta-function (time dependent Green function) [20]. Thus

$$\langle 0|T^{\alpha\beta}|0\rangle = i\hbar c \lim_{x\to x'}(3\partial^\alpha \partial'^\beta - 2g^{\alpha\beta}\partial_\nu \partial'^\nu)G(x,x') \qquad (10)$$

This means that to calculate the electromagnetic Casimir effect, it is enough to find the appropriate Green function corresponding to the geometry of the problem and then use the above relation for the vacuum to vacuum expectation value of the electromagnetic field stress tensor from which one can simply compute the desired Casimir force/pressure. The relation (10) is a general covariant formula which can be applied to different problems with different boundaries in flat (Minkowskian) space-time; this is because the metric tensor $g^{\mu\nu}$ with which we have worked is the Minkowskian metric tensor $g^{\mu\nu} = diag(1,-1,-1,-1)$. Of course, using quantum field theory in curved space-time, the method used here can be generalized for application in curved space-time geometries.

About the covariance of the formulation of the Casimir effect, we should mention that when a measurement is to be done, an experimentalist chooses a specific (Lab.) frame and then performs the experi-ment. In other words, to measure a non-covariant quantity for a particular problem (e.g. energy/force) with its special geometry and thus with its particular Green function, choosing an appropriate (specific) frame and working with non-covariant quantities (e.g. the distance between two plates) is unavoidable.

**Time dependent Green function (propagator) for a conducting spherical shell**

    A. Inside the Shell

Consider the following Fourier transform of the Green function (with Dirichlet boundary condition) for a conducting spherical shell of radius $a$

$$G(\vec{x},t,\vec{x}',t') = \frac{1}{2\pi}\int_0^\infty e^{-i\omega(t-t')}d\omega \sum_{l=0}^\infty \sum_{m=-l}^{l} g_l(r,r')Y_{lm}(\theta,\varphi)Y_{lm}^*(\theta',\varphi') \qquad (11)$$

in which $Y_{lm}$ s are spherical harmonics and $g_l(r,r')$ satisfies

$$\frac{1}{r}\frac{d^2}{dr^2}rg_l(r,r') - \frac{l(l+1)}{r^2}g_l(r,r') + \frac{\omega^2}{c^2}g_l(r,r') = \frac{1}{cr^2}\delta(r-r') \qquad (12)$$

The homogeneous form of the above equation is the famous spherical Bessel's differential equation; considering the symmetry of $g(r,r')$ under interchange of $r$ and $r'$, and the condition $g(r,r') \to 0$ when $r \to a$, and the fact that $g(r,r')$ should be finite at the origin, it is found [32]

$$g_l(r,r') = \frac{i\omega}{c^2} \frac{j_l\left(\frac{\omega}{c} r_<\right)}{j_l\left(\frac{\omega}{c} a\right)} \left[ h_l^{(1)}\left(\frac{\omega}{c} a\right) j_l\left(\frac{\omega}{c} r_>\right) - h_l^{(1)}\left(\frac{\omega}{c} r_>\right) j_l\left(\frac{\omega}{c} a\right) \right] \quad (13)$$

where $j_l\left(\frac{\omega}{c} r\right)$ and $h_l^{(1)}\left(\frac{\omega}{c} r\right)$ are spherical Bessel and Hankel functions of the first kind respectively [21] and $r_<(r_>)$ is the smaller (larger) values of $r$ and $r'$.

Therefore, the proper Green function for the inside of the shell has the following form:

$$G(\vec{x},t,\vec{x}',t')_{in} = \int_0^\infty \frac{i\omega}{2\pi c^2} \sum_{l=0}^\infty \sum_{m=-l}^l \frac{j_l\left(\frac{\omega}{c} r_<\right)}{j_l\left(\frac{\omega}{c} a\right)} \left[ h_l^{(1)}\left(\frac{\omega}{c} a\right) j_l\left(\frac{\omega}{c} r_>\right) - h_l^{(1)}\left(\frac{\omega}{c} r_>\right) j_l\left(\frac{\omega}{c} a\right) \right]$$

$$\times Y_{lm}(\theta,\varphi) Y_{lm}^*(\theta',\varphi') e^{-i\omega(t-t')} d\omega \quad (14)$$

### B. Outside the shell

Considering again the special symmetry of $g(r,r')$ under interchange of $r$ and $r'$, and the outside boundary conditions $g(r,r') \to 0$ when $r \to a$, and $g(r,r') \to$ free space Green function when $r \to \infty$, it is found:

$$G(\vec{x},t,\vec{x}',t')_{out} = \int_0^\infty \frac{i\omega}{2\pi c^2} \sum_{l=0}^\infty \sum_{m=-l}^l \frac{h_l^{(1)}\left(\frac{\omega}{c} r_>\right)}{h_l^{(1)}\left(\frac{\omega}{c} a\right)} \left[ h_l^{(1)}\left(\frac{\omega}{c} r_<\right) j_l\left(\frac{\omega}{c} a\right) - h_l^{(1)}\left(\frac{\omega}{c} a\right) j_l\left(\frac{\omega}{c} r_<\right) \right]$$

$$\times Y_{lm}(\theta,\varphi) Y_{lm}^*(\theta',\varphi') e^{-i\omega(t-t')} d\omega \quad (15)$$

**The electromagnetic Casimir effect for a real conducting spherical shell**

Insertion of (14) and (15) in the relation (10) gives us the $rr$ component of the energy-momentum tensors for the inside and outside of the shell as

$$\left\langle T^{rr}\big|_{in}\right\rangle = i\hbar c \lim \frac{\partial}{\partial x_i}\frac{\partial}{\partial x'_i} G(x,x')\Big|_{\substack{\varphi\to\varphi',\theta\to\theta' \\ t\to t', r\to r'=a}} = -\frac{i\hbar}{8\pi^2 a^2 c}\int_0^\infty \omega d\omega \sum_{l=0}^\infty (2l+1)\frac{j'_l\left(\frac{\omega}{c}a\right)}{j_l\left(\frac{\omega}{c}a\right)} \quad (16)$$

$$\left\langle T^{rr}\big|_{out}\right\rangle = i\hbar c \lim \frac{\partial}{\partial x_i}\frac{\partial}{\partial x'_i} G(x,x')\Big|_{\substack{\varphi\to\varphi',\theta\to\theta' \\ t\to t', r\to r'=a}} = \frac{i\hbar}{8\pi^2 a^2 c}\int_0^\infty \omega d\omega \sum_{l=0}^\infty (2l+1)\frac{h'^{(1)}_l\left(\frac{\omega}{c}a\right)}{h^{(1)}_l\left(\frac{\omega}{c}a\right)} \quad (17)$$

where $j'_l\left(\frac{\omega}{c}a\right) = \frac{c}{\omega}\frac{d}{dr}j_l\left(\frac{\omega}{c}r\right)\Big|_{r=a}$ and $h'^{(1)}_l\left(\frac{\omega}{c}a\right) = \frac{c}{\omega}\frac{d}{dr}h^{(1)}_l\left(\frac{\omega}{c}r\right)\Big|_{r=a}$.

Application of the complex frequency rotation ($\omega \to i\omega$) leads to:

$$\left\langle T^{rr}\big|_{in}\right\rangle = -\frac{\hbar}{8\pi^2 a^2 c}\int_0^\infty \omega d\omega \sum_{l=0}^\infty (2l+1)\frac{i'_l\left(\frac{\omega}{c}a\right)}{i_l\left(\frac{\omega}{c}a\right)} \quad (18)$$

$$\left\langle T^{rr}\big|_{out}\right\rangle = \frac{\hbar}{8\pi^2 a^2 c}\int_0^\infty \omega d\omega \sum_{l=0}^\infty (2l+1)\frac{k'_l\left(\frac{\omega}{c}a\right)}{k_l\left(\frac{\omega}{c}a\right)} \quad (19)$$

in them $i_l$ and $k_l$ are modified spherical Bessel functions [37].

The vacuum state is the ground state in which the orbital quantum number $l$ should take its minimum value (see the **Appendix**); so, although $\omega$ takes a wide variety of possible values, $l$ should be set equal to zero for the vacuum to vacuum energy-momentum tensor expectation value. Therefore, equations (18) and (19) reduce to:

$$\left\langle T^{rr}\big|_{in}\right\rangle = -\frac{\hbar}{8\pi^2 a^2 c}\int_0^\infty \omega d\omega \frac{i'_0\left(\frac{\omega}{c}a\right)}{i_0\left(\frac{\omega}{c}a\right)} \quad (20)$$

$$\left\langle T^{rr}\big|_{out}\right\rangle = \frac{\hbar}{8\pi^2 a^2 c}\int_0^\infty \omega d\omega \frac{k'_0\left(\frac{\omega}{c}a\right)}{k_0\left(\frac{\omega}{c}a\right)} \quad (21)$$

To calculate the desired Casimir force, we need to find the following subtraction

$$\langle T^{rr}|_{in}\rangle - \langle T^{rr}|_{out}\rangle = -\frac{\hbar}{8\pi^2 a^2 c} \int_0^\infty \omega \left( \frac{i_0'\left(\frac{\omega}{c}a\right)}{i_0\left(\frac{\omega}{c}a\right)} + \frac{k_0'\left(\frac{\omega}{c}a\right)}{k_0\left(\frac{\omega}{c}a\right)} \right) d\omega \qquad (22)$$

The above integral expression has an infinite value; we have reached an irregular result that must be renormalized in some way. We regularize the above relation with plasma frequency cut-off integration. As was explained in the introduction, for frequencies higher than the plasma frequency, the surface of the shell behaves transparently and the virtual photons of the vacuum state do not "see" any boundary; this means that for frequencies $\omega > \omega_p$, there is no difference between "in" and "out" and the upper bound of the above integral formula should be set equal to the cut-off value $\omega_p$. This fact that a real material cannot constrain modes of the field with frequencies much higher cut-off frequency has been already considered for the Casimir scalar field energy with the geometry of plate(s) and sphere [38]. Therefore, we can write the following regularized form for the relation (37):

$$\left(\langle T^{rr}|_{in}\rangle - \langle T^{rr}|_{out}\rangle\right)_{regularized} = -\frac{\hbar}{8\pi^2 a^2 c} \int_0^{\omega_p} \omega \left( \frac{i_0'\left(\frac{\omega}{c}a\right)}{i_0\left(\frac{\omega}{c}a\right)} + \frac{k_0'\left(\frac{\omega}{c}a\right)}{k_0\left(\frac{\omega}{c}a\right)} \right) d\omega \qquad (23)$$

Using $\left(i_0(x) = \frac{\sinh x}{x}\right)$ and $\left(i_0(x) = \frac{\sinh x}{x}\right)$ and knowing that for real (e.g. golden or silver) spherical shells of radii at the scales of less than $10^{-7} m$ (a distance scale at which the Casimir effect is noticeable), $(\frac{\omega_p}{c}a)$ is less than 1 (see **Table 1**), we simply find:

$$\int_0^{\omega_p} \omega \left( \frac{i_0'\left(\frac{\omega}{c}a\right)}{i_0\left(\frac{\omega}{c}a\right)} + \frac{k_0'\left(\frac{\omega}{c}a\right)}{k_0\left(\frac{\omega}{c}a\right)} \right) d\omega \approx -\int_0^{\omega_p} \omega \frac{1}{\left(\frac{\omega}{c}a\right)} d\omega = -\left(\frac{\omega_p}{a}c\right) \qquad (24)$$

Using (23) and (24):

$$\left(\langle T^{rr}|_{in}\rangle - \langle T^{rr}|_{out}\rangle\right)_{renormalizd} \approx \frac{\hbar \omega_p}{8\pi^2 a^3} \qquad (25)$$

Knowing that:

$$f = \int_0^{2\pi} a^2 d\varphi \int_{-1}^{1} d(\cos\theta) \left( \langle T^{rr} |_{in} \rangle - \langle T^{rr} |_{out} \rangle \right)_{renormalizd}$$
$$= 4\pi a^2 \left( \langle T^{rr} |_{in} \rangle - \langle T^{rr} |_{out} \rangle \right)_{renormalizd} \quad (26)$$

The Casimir force is found as:

$$f \approx \frac{\hbar \omega_p}{2\pi a} \quad (27)$$

Comparing to the well-known results (e.g. the result by Boyer [5–8]), we expect a repulsive force with an inverse square dependence on the shell radius ( $f_{Boyer} \approx 0.046(\frac{\hbar c}{a^2})$ ); but why has an inverse linear form been found? We should be careful that our computation here is based on the smallness of the dimensionless parameter ($\frac{\omega}{c}a$); this is a scale at which the radius of the sphere is small compared to the wavelength associated with the frequency at which the shell material becomes transparent (non-retarded limit). The perfect conductor limit, the limit found by Boyer, formally corresponds to a retarded (long distance) limit and has a different (inverse square) functional form. It is a well-known result in Casimir force calculations that the non-retarded forces differ by one inverse power from the result found in the retarded limit.

**Table 1**. The numerical values of the Casimir force in nanoNewtons.

| Material | $\omega_p$ ( $rad/sec$ ) | $a$ (m) | $\left(\frac{\omega_p}{c}a\right)$ | $f$ (nanoNewton) |
|---|---|---|---|---|
| Gold | $1.37 \times 10^{16}$ | $10^{-8}$ | 0.4567 | 0.0230 |
| Gold | $1.37 \times 10^{16}$ | $10^{-9}$ | 0.0457 | 0.2300 |
| Silver | $9.65 \times 10^{14}$ | $10^{-8}$ | 0.0322 | 0.0016 |
| Silver | $9.65 \times 10^{14}$ | $10^{-9}$ | 0.0032 | 0.0162 |

**Conclusion**

Using covariant formulation of the electromagnetic Casimir effect with a regularization method based on plasma frequency cut-off integration, the Casimir force for a real conducting spherical boundary was computed. The numerical results for real cases of golden and silver shells have been shown in **Table 1**. The resulting force is repulsive with an inverse linear dependence on the shell radius. The numerical values of the Casimir force (in nanoNewtons) have been computed in the last column of the table. Higher conductivity and a smaller radius lead to a greater value of the Casimir force. This can be understood qualitatively based on the conceptual argument about the pressure made by virtual photons confined into the shell region. As is clear from the data in Table 1, the numerical results directly depend on the cut-off values (material kinds); this is an unavoidable property of the Casimir energy for real materials [38].

In this paper, we have only computed the electro-magnetic Dirichlet Casimir force without considering the Neumann problem. In the already known non-covariant formalisms, one considers the Neumann problem parallel to the Dirichlet problem and then adds the results. This is because, in non-covariant formalisms, the electric and magnetic field energies separately contribute to the total energy of the electro-magnetic field and the fields are (often) divided into TE and TM modes with Dirichlet and Neumann boundary conditions; but, here in the covariant approach, the Dirichlet and Neumann boundary value problems are two different/separate problems. In the covariant quantization of the electromagnetic field, one quantizes $A^\mu$ and doesn't work with the non-covariant objects $E_i$ and $B_i$. Although $E_i$ and $B_i$ have the main physical role of the electromagnetic fields in classical non-covariant arguments, we have quantized the covariant four potential $A^\mu$ (as a basis for the electromagnetic field tensor $F^{\mu\nu}$) similar to the standard canonical quantum theory of the electromagnetic field. Since the potential $A^\mu$ satisfies either the Dirichlet or Neumann boundary separately, we have studied the Dirichlet problem independently.

*** *Fermi Lagrangian density is simply found from the well-known Lagrangian density $L = -\frac{1}{4} F_{\mu\nu} F^{\mu\nu}$ by choosing the covariant Lorentz gauge. It has no difficulty in the introduction of the zero component of momentum (conjugate) field* [36]. *Although working in Lorentz gauge has other difficulties that can be removed with the formalism introduced by Gupta and Bleuler* [39-40], *our physical results here corresponding to the Casimir effect are independent of choosing any particular gauge.*

**Appendix**

Based on the uncertainty principle, the virtual particles are called virtual because they are created and annihilated in a distance and with a lifetime less than the fundamental minimum observable values known from the relations $\Delta p \Delta x \geq \frac{\hbar}{2}, \Delta E \Delta t \geq \frac{\hbar}{2}$. Both these relations reason on this fact that the maximum angular momentum value of a virtual particle is bounded in the range of the first order of magnitude of the fundamental constant $\hbar$ and thus the eigenvalue $l(l+1)\hbar^2$ corresponding to the operator $\hat{L}^2$ for virtual particles cannot be more than $\hbar^2$ which is impossible unless the orbital quantum number $l$ is set equal to zero. More technically, let consider a virtual particle of mass $m$ (paying enough attention to this fact known from quantum field theory that the virtual photons have mass and their velocity is less than the free value $c$), which is in the existence in a distance $x$ (where $x$ is less than the sphere diameter $2a$ for inner particles and has any positive value for outer ones in the sphere problem here is under study), with a lifetime $\tau$ which should satisfy the following momentum-distance and energy-time relations:

$$\begin{cases} px < \dfrac{\hbar}{2} \\ E\tau < \dfrac{\hbar}{2} \end{cases} \quad \text{(A-1)}.$$

For a virtual particle with the velocity $v$, the above relation can be written as:

$$\begin{cases} mvx < \dfrac{\hbar}{2} \\ E(\dfrac{x}{v}) < \dfrac{\hbar}{2} \end{cases} \qquad (A\text{-}2).$$

As we know, the orbital angular momentum contribution to the energy appears as $\dfrac{l(l+1)\hbar^2}{2mx^2}$; this is clearly less than the energy value $E$ because the only other term contributing to the virtual particle energy is its kinetic term corresponding to its momentum $p$ which has a positive value. Therefore, using (A-2):

$$\dfrac{l(l+1)\hbar^2}{2mx^2} \leq E \rightarrow \dfrac{l(l+1)\hbar^2}{2mx^2}(\dfrac{x}{v}) < \dfrac{\hbar}{2} \rightarrow l(l+1)\hbar < mvx \rightarrow l(l+1) < \dfrac{1}{2} \rightarrow l = 0 \qquad (A\text{-}3).$$